\documentstyle[12pt,epsfig]{article}
\newcommand{\bm}{\bibitem}

\def\be {\begin{equation}}
\def\ee {\end{equation}}
\def\bea {\begin{eqnarray}}
\def\eea {\end{eqnarray}}
\def\nn {\nonumber}

\newcommand {\om}{{\omega}}

\begin{document}

\thispagestyle{empty}
\vskip 15pt

\begin{center}
{\Large {\bf Nuclear modification factor in an anisotropic 
{\em Quark-Gluon-Plasma}}}
\renewcommand{\thefootnote}{\alph{footnote}}

\hspace*{\fill}

\hspace*{\fill}

{ \tt{
Mahatsab Mandal\footnote{E-mail address:
mahatsab.mandal@saha.ac.in}
Lusaka Bhattacharya\footnote{E-mail address:
lusaka.bhattacharya@saha.ac.in}
and
Pradip Roy\footnote{E-mail address:
pradipk.roy@saha.ac.in}
}}\\

\small {\em Saha Institute of Nuclear Physics \\
1/AF Bidhannagar, Kolkata - 700064, INDIA}
\\

\vskip 40pt

{\bf ABSTRACT}
\end{center}

\vskip 0.5cm

We calculate the nuclear modification factor ($R_{AA}$) of light
hadrons by taking into account the initial state momentum
anisotropy of the quark gluon plasma (QGP) expected to be formed in
relativistic heavy ion collisions. 
Such an anisotropy can result from the 
initial rapid longitudinal expansion of the matter.
A phenomenological model for the space time evolution of the
anisotropic QGP is used to obtain the time dependence of the
anisotropy parameter $\xi$ and the hard momentum scale, $p_{\rm hard}$. 
The result is then compared with the PHENIX experimental data to
constrain the isotropization time scale, $\tau_{\rm iso}$ for fixed initial
conditions (FIC).
It is shown that the extracted value of $\tau_{\rm iso}$ lies in the
range $0.5 \leq \tau_{\rm iso} \leq 1.5$.
However, using fixed final multiplicity (FFM) condition does
not lead to any firm conclusion about the extraction of the
isotropization time.
The present calculation is also extended to contrast with
the recent measurement of nuclear modification factor by ALICE
collaboration at $\sqrt{s}=2.76$ TeV. It is argued that in the present
approach, the extraction
of $\tau_{\rm iso}$ at this energy is uncertain and, therfore, refinement of the model 
is necessary.
 The sensitivity of
the results on the initial conditions has been discussed. We also present
the nuclear modification factor at LHC energies with $\sqrt{s} = 5.5$ TeV.

\vskip 30pt

\section{Introduction}

The partonic energy loss in a QCD plasma has received
significant attention in recent years.  Experimentally, the partonic energy
loss can be probed by measuring the high $p_T$ hadrons emanating from
ultra-relativistic heavy ion collisions. This idea was first
proposed by Bjorken \cite{bjorken} where `ionization loss' of
the quarks and gluons in a QCD plasma was estimated. In fact, the `stopping
power' ($dE/dx$) of the plasma is proportional
to $\sqrt{\epsilon}$, where, $\epsilon$
is the energy density of the partonic medium. Therefore, by measuring
various high $p_T$ observables one can probe the initial parton density
\cite{bjorken}.

Hard partons, injected into hot QCD medium, can dissipate energy in two
ways, {\em viz.}, by two body collisions or {\em via} the bremsstrahlung
emission of gluons, commonly referred to as collisional and radiative loss
respectively.  For electromagnetic processes, it is well known that at large
energies, radiative losses are much higher than the collisional loss.

The possibility of QGP formation at RHIC 
experiment, with initial density of $5~GeV/fm^{3}$   
is supported by the observation of high $p_T$ hadron suppression
(jet-quenching) in the central Au-Au collisions compared to the 
binary-scaled hadron-hadron collisions~\cite{jetquen}. 
The phenomena of jet-quenching has been investigated
by various authors~\cite{jetquen}.
Apart form jet-quenching, 
several possible probes have been studied in order to characterize the 
properties of QGP.

However, many properties of QGP are still poorly understood. The most 
debated question is whether the matter formed in the relativistic heavy ion
collisions is in thermal equilibrium or not. The measurement of elliptic
flow parameter and its theoretical explanation suggest that the matter
quickly comes into thermal equilibrium 
(with $\tau_{\rm therm} < $ $1$ fm/c, where $\tau_{therm}$ is the time of 
thermalization)~\cite{PRC75_ref1}.
As for example, 
one of the major difficulty is to measure the thermalization 
($\tau_{therm}$) and isotropization ($\tau_{iso}$) time of the QGP. On 
the one hand, the success of ideal hydrodynamic fits to experimental 
data~\cite{PRC75_ref1} implies rapid thermalization of the bulk matter 
created at RHIC. 
On the contrary, perturbative estimation suggests relatively
slower thermalization of
QGP~\cite{PRC75_ref2}. However, recent hydrodynamical 
studies~\cite{0805.4552_ref4} have shown that 
due to the poor knowledge of the initial conditions there is a
sizable amount of uncertainty in the estimate of thermalization or 
isotropization time. It is suggested that (momentum) anisotropy driven
plasma instabilities may speed up the process of 
isotropization~\cite{plb1181993}, in that case one is allowed to use 
hydrodynamics for the evolution of the matter. However, instability-driven 
isotropization is not yet proved at RHIC and LHC energies. 

In absence of a theoretical proof favoring the rapid
thermalization and the uncertainties in the hydrodynamical fits of
experimental data, it is very hard to assume hydrodynamical
behavior of the system from the very beginning. Therefore, 
it has been suggested to look for some observables which are sensitive to the
early time after the collision. For example, jet-quenching vis-a-vis energy
loss of partons could be an observable where the initial state
momentum anisotropy can play important role. 
It is to be noted that the calculations of energy loss in Ref.~\cite{jetquen}
have been performed in isotropic QGP which is true immediately
after its formation~\cite{5of4552}. However, subsequent rapid
expansion of the matter along the beam direction causes faster cooling in
the longitudinal direction than in the transverse
direction~\cite{PRC75_ref2}. As a result, the system becomes anisotropic
with $\langle{p_L}^2\rangle << \langle{p_T}^2\rangle$ in the local rest frame.
At some later time when the effect of parton interaction rate
overcomes the plasma expansion rate, the system returns to the
isotropic state again and remains isotropic for the rest of the period.
Thus, during the early stage the plasma remains anisotropic and any calculation
of energy loss should, in principle, include this aspect.
The collisional energy loss in anisotropic media for heavy fermions has been 
calculated in Refs.~\cite{strick1,strick}. 
In these calculations it is found that the deviations from the isotropic
results are of the order of 10\% for $\xi=1$ ($\xi$ is the
anisotropy parameter)  and of the order of
20\% for $\xi=10$. It is observed that the collisional energy loss
varies with the angle of propagation by upto 50\%.

Radiative energy loss
in anisotropic QGP has recently been calculated in Ref.~\cite{prabhee} in
first order opacity expansion where the scatterers are treated as static.
It should be noted that the energy loss of a parton in anisotropic media
depends both on the anisotropy parameter and the direction of propagation
with respect to the anisotropy axis ${\hat n}$. In that case, general 
expression for the two-body potential should be used~\cite{prabhee}. 
Few comments about the calculation of radiative energy loss
are in order here.
First of all we have considered scattering from
static charges in which case only longitudinal gauge bosons
are exchanged rendering the screened potential stable. 
The
effect of anisotropy would be significantly different if one
considers the recoil of the scattering centre.
In case of moving scatterer, transverse (magnetic) gluon exchange 
results plasma instabilities which has to be taken into account.
It is worthwhile to mention that the 
growth of unstable modes in anisotropic media greatly influences
the transport co-efficients and hence the energy loss.
This is recently demonstrated in Ref.~\cite{abhijit}. 
Considering a two-stream plasma
the authors in ~\cite{abhijit} show that the momentum broadening
grows exponentially with time
as the spontaneously growing fields exert an exponentially growing
influence on the propagating parton. In an evolving plasma this aspect
is an important component without which the results remain somewhat
like a zeroth order approximation.
However, in the present work, we do not include this effect and
assume that the fragmenting partons propagate in the direction of anisotropy.

For a parton propagating in the direction of anisotropy it is found that 
that the fractional energy loss increases by a factor
of 1.5 - 2 depending upon the anisotropy parameter $\xi$. In the present work,
we shall apply this formalism to calculate the nuclear modification factor
for light hadrons. A phenomenological model for the space-time
evolution will be used for the time evolution of $\xi$ and $p_{\rm hard}$.
Since the role of collisional energy loss in the context of RHIC data
is not settled yet, we shall not include this process in the present work.

The plan of the paper is the following. In section 2 we briefly
recall the necessary ingredients to calculate radiative energy loss in 
anisotropic media. Then we discuss how this can be implemented to
calculate $R_{\rm AA}$ along with space-time model for
anisotropic media.
Section 3 will be devoted
to discuss the results. Finally, we conclude in section 4.

\section{Formalism}   
\subsection{Radiative energy loss}

In this section, we briefly mention the formalism of the radiative
energy loss in an infinitely extended anisotropic plasma (see 
Ref.~\cite{prabhee} for further details). 

We assume that an on-shell quark produced in the remote past
is propagating through an infinite QCD medium that consists of randomly
distributed static scattering centers which provides
a color-screened Yukawa
potential originally developed for the isotropic QCD medium given 
by~\cite{gulwangnpb}
\bea
V_n&=&V(q_n)e^{i{\vec q}_n \cdot {\vec x}_n}\nn\\  
   &=&2\pi \delta(q^0)v(q_n)e^{-i{\vec q}_n \cdot {\vec x}_n}T_{a_n}(R)
\otimes T_{a_n}(n).
\eea
with $v({\vec q}_n)=4\pi\alpha_s/({\vec q}_n^2+\mu^2)$, where $\mu$ is the Debye
mass. 
$x_n$ is the location of the $n$th scattering centre, $T$ (summed
over $a_n$) denotes the colour matrices of the parton and the scattering
centre. It is to be noted that the potential has been derived by using
hard thermal loop (HTL) propagator in QGP medium. In a plasma with 
momentum anisotropy the two body interaction, as expected, becomes direction 
dependent. In addition, it depends on the anisotropy parameter $\xi$ and
the hard momentum scale $p_{\rm hard}$ in the following way: 
\be
f(\vec p) = \frac{1}{e^{p/p_{\rm hard}\sqrt{1+\xi ({\hat p}\cdot n)^2}}\pm 1} 
\ee 
where, the parameter $\xi$ is the degree of anisotropy 
parameter ($ -1 < \xi < \infty $) and  is given by
$\xi=\langle p_T^2 \rangle/(2 \langle p_z^2 \rangle)-1$. 
It is to be noted 
that $\xi$ can also be related to the shear viscosity~\cite{asakawa}.

To calculate the two body potential appearing in Eq.(1) one starts with
the retarded gluon self energy expressed as~\cite{iancu}
\be
\Pi^{\mu \nu}(P) = g^2\,\int\,\frac{d^3k}{(2\pi)^3}v^{\mu}\,
\frac{\partial f({\vec k})}{\partial K^\beta}
\left(g^{\nu \beta}-\frac{v^{\nu} P^\beta}{P\cdot v+i\epsilon}\right)
\label{se}
\ee
We have adopted the following notation for four vectors:
$P^{\mu} = (p_0,{\vec p}) = (p_0,{\bf p},p_z)$, i. e. ${\vec p}$
(with an explicit vector superscript) describes a three-vector while
${\bf p}$ denotes the two-vector transverse to the $z$-direction.

To include the local anisotropy in the plasma, one has to 
calculate the gluon polarization tensor incorporating anisotropic distribution
functions of the medium. This subsequently can be used to construct
HTL corrected gluon propagator which, in general,
assumes very complicated from. Such an HTL propagator was first
derived in \cite{Romash} in time-axial gauge. Similar propagator
has also been constructed in \cite{dumitru08} to derive the heavy-quark
potential in an anisotropic plasma, which, as we know, is given by the 
Fourier transform of the propagator in the static limit.

The self-energy, apart from momentum $P^\mu$, also depends on a
fixed anisotropy vector $n^{\mu} (=(1,{\vec n}))$ and $\Pi^{\mu \nu}$
can be cast in a suitable tensorial basis appropriate
for anisotropic plasma in a co-variant gauge in the
following way~\cite{dumitru08}:
\be
\Pi^{\mu \nu} = \alpha\,A^{\mu \nu}+\beta\,B^{\mu \nu}+\gamma\,C^{\mu \nu}
+\delta\,D^{\mu \nu}
\ee
where the basis tensors are constructed out of $p^\mu$, $n^\mu$ and the
4-velocity of the heat bath $u^{\mu}$. The detailed expressions for
the quantities those appear in Eq.(4) can be found in 
Ref.~\cite{Romash,dumitru08}.
The anisotropy enters through the distribution function given earlier 
(see Eq.(2)).

Since the self-energy is symmetric and transverse, all the components
are not independent. After change of variables ($p^\prime = 
{\vec p}^2[1+\xi({\hat {\bf p}}
\cdot {\vec n})^2]$) the spatial components can be written as
\be
\Pi^{i j} = \mu^2\,\int\,\frac{d\Omega}{4\pi}\,v^i
\frac{v^l+\xi({\vec v}\cdot {\vec n})n^l}{1+\xi({\vec v}\cdot {\vec n})^2}
\left(\delta^{j l}+\frac{v^j p^l}{P\cdot v +i\epsilon}\right)
\ee
Now $\alpha, \beta, \gamma$ and $\delta$ are determined by the following
contractions:
\bea
p^i\,\Pi^{i j}\,p^j& =& {\vec p}^2\beta\nonumber\\
A^{i l}\,n^l\,\Pi^{i j}\,p^j& =& ({\vec p}^2-(n\cdot P)^2)\delta\nonumber\\
A^{i l}\,n^l\,\Pi^{i j}\,A^{j k}\,n^k& =&
\frac{{\vec p}^2-(n\cdot P)^2}{{\vec p}^2}(\alpha+\gamma)\nonumber\\
{\rm Tr}\Pi^{i j}& =& 2\alpha+\beta+\gamma
\eea
where the expressions for $\alpha, \beta, \gamma$ and $\delta$ are given
in Ref.~\cite{Romash}.

After knowing the gluon HTL self-energy in anisotropic
media the propagator can be calculated after some cumbersome
algebra~\cite{dumitru08,baier}:
\bea
\Delta^{\mu\nu}&&=\frac{1}{(P^2-\alpha)}\big [A^{\mu\nu}-C^{\mu\nu} \big ]\nn\\
&& + \Delta_G \Big [(P^2-\alpha-\gamma)\frac{\om^4}{P^4}B^{\mu\nu}+
(\om^2-\beta) C^{\mu\nu}+\delta \frac{\om^2}{P^2}D^{\mu\nu} \Big]
-\frac{\lambda}{P^4}P^\mu P^\nu
\eea
where
\bea
\Delta_G^{-1}=(P^2-\alpha-\gamma) (\om^2-\beta) -\delta^2 [P^2-(n\cdot P)^2]
\eea
It is well known that the momentum space potential can be obtained from 
the static gluon propagator in the following way:
\bea
v({\bf q},\xi)\equiv v({\bf q},q_z=0,\xi)&=&g^2\,\Delta^{00}(\omega=0,{\bf q},q_z=0,\xi)
\nonumber\\
&=&g^2\,\frac{{\vec q}^2+m_\alpha^2+m_\gamma^2}
{({\vec q}^2+m_\alpha^2+m_\gamma^2)({\vec q}^2+m_\beta^2)-m_{\delta}^4}
\label{pot1}
\eea
where, in general, the expressions for $m_\alpha^2, m_\beta^2, m_\gamma^2 $ 
and $m_\delta^2$ are lengthy~\cite{dumitru08}. For $q_z=0$ (which is
the case here) these simplify to~\cite{prabhee}
\bea
m_\alpha^2&=&0\nonumber\\
m_\beta^2&=& \mu^2\,R(\xi)\nonumber\\
m_\gamma^2&=&-\mu^2\left[\frac{1}{1+\xi}-R(\xi)\right]\nonumber\\
m_\delta^2&=&0
\eea
for a parton propagating in the anisotropy direction.
Here $\mu$ is the Debye mass given by $\mu^2 = g^2\,p_{\rm hard}^2\,(1+N_F/6)$,
where $N_F$ is the number of flavours.
In such case the two-body potential in anisotropic media simplifies 
to~\cite{prabhee}

\be
v({\bf q},\xi)=\frac{4 \pi \alpha_s}{{\bf q}^2+R(\xi)\mu^2} 
\label{pot2}
\ee
with
\be
R(\xi)=\frac{1}{2} \left [\frac{1}{1+\xi} + \frac{\tan ^{-1}\sqrt{\xi}}
{\sqrt{\xi}}\right ]
\ee

Now the parton scatters with one of the
colour centre with the momentum $Q = (0,{\bf q},q_z)$ and radiates a
gluon with momentum $K = (\omega,{\bf k},k_z)$. The method for calculating
the amplitudes
is discussed in Refs.~\cite{magdalenaprc,mplb,mnpa} and we shall ,for
the sake of brevity, quote the main
results only.
The
quark energy loss is calculated by folding the rate of gluon radiation
($\Gamma (E)$) with the gluon energy by assuming $\omega+q_0 \approx
\omega $. In this approximation one finds~\cite{magdalenaprc},

\bea
\frac{dE}{dL}&=
&\frac{E}{D_R} \int x dx \frac{d\Gamma}{dx}
\eea
Here $D_R$ is defined as $[t_a,t_c][t_c,t_a]=C_2(G)C_RD_R$,
where $C_2(G)=3$, $D_R=3$ and $[t_a,t_c]$ is a color commutator
(see ~\cite{magdalenaprc} for details). $x$ is the longitudinal
momentum fraction of the quark carried away by the emitted gluon.

Now in anisotropic media we have~\cite{prabhee},

\bea
x \frac{d\Gamma}{dx}&=&\frac{C_R\alpha_s}{\pi} \frac{L}{\lambda}
\int \frac{d^2{\bf k}}{\pi}\frac{d^2{\bf q}}{\pi} |v({\bf q},\xi)|^2 
\frac{\mu^2}{16\pi^2\alpha_s^2}
\left [\frac{{\bf {k+q}}}{({\bf{k+q}})^2+\chi^2}-\frac{{\bf k}}{{\bf k}^2+\chi} \right ]^2
\eea

In the last expression, $v({\bf q},\xi)$ is the two body quark-quark potential 
given by Eq.(\ref{pot2}) and $\chi = m_q^2x^2+m_g^2$, where $m_g^2=\mu^2/2$ 
and $m_q^2 = \mu^2/6$.

For arbitrary $\xi$ the radiative energy loss can be written as~\cite{prabhee}
\bea
\frac{\Delta E}{E}&=&\frac{C_R\alpha_s}{\pi^2} \frac{L}{\lambda}
\int dx d^2{\bf q}| \frac{\mu^2}{({\bf q}^2+R(\xi)\mu^2)^2}
\left [\frac{}{}   
-\frac{1}{2}\right.\nn\\
&&\left.-\frac{k_m^2}{k_m^2+\chi} + 
\frac{{\bf q}^2-k_m^2+\chi}{2 \sqrt{{\bf q}^4+ 2 {\bf q}^2 (\chi-k_m^2)}+(k_m^2+\chi)^2
}+\right.\nn\\
&&
\left.
\frac{{\bf q}^2+ 2 \chi} {{\bf q}^2 \sqrt{1+ \frac{4\chi}{{\bf q}^2} }}\ln
\left ( 
\frac{k_m^2+\chi}{\chi} 
\frac{({\bf q}^2+3 \chi) + \sqrt{ 1+ 
\frac{4\chi}{{\bf q}^2} } ({\bf q}^2+ \chi)}{({\bf q}^2-k_m^2+3 \chi) 
+  \sqrt{1+ \frac{4\chi}{{\bf q}^2} } \sqrt{{\bf q}^4+ 2 {\bf q}^2 (\chi-k_m^2)}+(k_m^2+\chi)^2 }
\right)
\right ]\nn\\
\label{debye}
\eea
In the above expression, $\lambda$ denotes the average mean free path
of the quark given by
\begin{equation}
\frac{1}{\lambda} = \frac{1}{\lambda_g}+ \frac{1}{\lambda_q}
\end{equation}
which in this case would be $\xi$ dependent. 
Note that $\lambda_g$ and $\lambda_q$ correspond to the contributions to
the mean free path of the propagating quark coming from
$q$-$g$ and $q$-$q$ scatterings.

Explicitly with
Eq.(\ref{pot2}) we have,
\begin{equation}
\lambda_i^{-1} = \frac{C_R C_2(i) \rho_i}{d_A}\,
\int\,d^2{\bf q}\,\frac{4\alpha_s^2}{({\bf q}^2+R(\xi)\mu^2)^2}.
\end{equation}
where $C_R=4/3$, $C_2(i)$ is the cashimir for $d_i$-dimensional
representation and $C_2(i)=(N_c^2-1)/(2N_c)$ for quark and
$C_2(i)=N_c$ for gluon scatterers. $d_A=N_c^2-1$ is the dimensionality of the
adjoint representation and $\rho_i$ is the density of the scatterers.
Using $\rho_i=\rho_i^{\rm iso}/\sqrt{1+\xi}$
we obtain
\begin{equation}
\frac{1}{\lambda} = \frac{18\alpha_s p_{\rm hard}\zeta(3)}{\pi^2\sqrt{1+\xi}} \frac{1}{R(\xi)}
\frac{1+N_F/6}{1+N_F/4}
\label{al1}
\end{equation}
For $\xi \rightarrow 0$ Eq.(\ref{al1}) reduces to well-known
results~\cite{magdalenaprc}
\begin{equation}
\frac{1}{\lambda} = \frac{18\alpha_s T\zeta(3)}{\pi^2}
\frac{1+N_F/6}{1+N_F/4}
\end{equation}
 

\subsection{Hadronic $p_T$ spectrum}
The high $p_T$ inclusive hadron spectrum in a heavy ion collision can be 
calculated in a pQCD-improved parton model.
There are various approaches about how to incorporate the energy loss in the
hadron production from jet fragmentation. The differential cross-section for
hadron production is~\cite{owens},
\begin{eqnarray}
E\frac{d\sigma}{d^3p} (AB\rightarrow {\rm jet}+X) & = &
K\,\sum_{abcd}\, \int dx_a dx_b
G_{a/h_A}(x_a,Q^2)
\,G_{b/h_B}(x_b,Q^2)
\nonumber\\
&\times&\frac{{\hat s}}{\pi}
\frac{d{\sigma}}
{d{\hat t}}(ab\rightarrow cd)\delta({\hat {s+t+u}}),
\label{hadpt}
\end{eqnarray}
Expressing the argument of the $\delta$-function in terms of $x_a$ and
$x_b$ and doing the $x_b$ integration we arrive at the final expression:
\begin{eqnarray}
E\frac{d\sigma}{d^3p} (AB\rightarrow {\rm jet}+X) & = &
K\,\sum_{abcd}\, \int_{x_{\rm amin}}^1 dx_a 
G_{a/h_A}(x_a,Q^2)
\,G_{b/h_B}(x_b,Q^2)
\nonumber\\
&\times&\frac{2}{\pi}\frac{x_a x_b}{2x_a-x_Te^y}
\frac{d{\sigma}}
{d{\hat t}}(ab\rightarrow cd)
\label{hadpt1}
\end{eqnarray}
where $x_b=(x_ax_Te^{-y})/(2x_a-x_Te^y)$, $x_T=2p_T/\sqrt{s}$ and 
$x_{\rm amin}=(x_Te^y)/(2-x_Te^{-y})$.
It should be noted that to obtain single particle inclusive
invariant cross-section in hadron-hadron collisions, the fragmentation
function $D_{h/c}(z,Q^2)$ must be included. 
Multiplying the result by the nuclear 
overlap function for a given centrality one can obtain the
$p_T$ distribution of hadrons in $A-A$ collisions.
However, the inclusion of 
jet-quenching as a final state effect in nucleus-nucleus collisions
, can be implemented in two ways:
(i) modifying
the fragmentation function~\cite{prc582321} and (ii) modifying the
partonic $p_T$ spectra~\cite{roy06} but keeping the fragmentation
function unchanged. We shall concentrate on the former approach, as it
is easier to include the final state effect in presence of anisotropy.
Now the energy loss of high energy quarks and gluons traveling through
dense colored plasma can measure the integrated density of the colored
particles. This non-Abelian energy loss is a function of parton opacity
$L/\lambda$. We use the expression given by Eq.(\ref{debye})
which is derived to first order in opacity.

Now in order to obtain the hadronic $p_T$ spectrum in $A-A$ collisions we
modify the fragmentation function to obtain an effective 
fragmentation function as follows:
\be
D_{h/c}(z,Q^2)=\frac{z^\ast}{z}D_{h/c}(z^\ast,Q^2)
\label{modfrag}
\ee
where, $z^\ast = z/(1-\Delta E/E)$ is the modified momentum fraction.
Also in order to take into account the jet production geometry
we assume that all the jets are not produced at the same
point and the path length traversed by these partons before
fragmentation are not the same.
It is also assumed that
the jet initially produced at $(r,\phi)$
leaves the plasma after a proper time ($t_L$) or equivalently after
traversing a distance $L$ (for light quarks $t_L \sim L$) given by
$L(r,\phi)=\sqrt{R_T^2-r^2\sin{\phi}^2}-R_T\cos{\phi}$, where
$R_T$ is the transverse dimension of the system.
Now the hadron $p_T$ spectra depends on the path length the
initial parton must travel and the temperature profile along that
path. As mentioned this is not the same for the all jets as it depends on
the location where the jet is produced. Therefore we have to convolute
the resulting expression over all transverse positions.
Since the number of jets produced at $\vec r$ is proportional to the
number of binary collisions, the probability is proportional to
the product of thickness functions :

\be
{\cal P}({\vec r}) \propto T_A({\vec r})\,T_B({\vec r})
\ee
For a hard sphere ${\cal P}(r)$ is given by 
\be
{\cal P}(r) = \frac{2}{\pi R_T^2}\left(1-\frac{r^2}{R_T^2}\right)\theta(R_T-r)
\ee
where $\int d^2r\,{\cal P}(r)=1$.
Using all these and noting that the path length is not a measurable quantity
we obtain the $p_T$ spectra of hadrons as follows:
  
\bea
\frac{dN^{\pi^0(\eta)}}{d^2p_Tdy} &=& \sum_f\,\int\,d^2r{\cal P}(r)\,
\int_{t_i}^{t_L}\,\frac{dt}{t_L-t_i}\int\,
\frac{dz}{z^2}\nonumber\\
&&\times\,D_{\pi^0(\eta)/f}(z,Q^2)|_{z=p_T/p_T^f}
\,E\frac{dN}{d^3p^f},
\eea
The quantity $E\frac{dN}{d^3p^f}$ is the initial momentum distribution
of jets and can be computed using LO-pQCD (see Eq.(\ref{hadpt1})).
For the fragmentation function Eq.(\ref{modfrag}) has been used.
We use the average value of distance traversed by the partons,
$\langle L \rangle $ given by
\be
\langle L \rangle=
\frac{\int_0^{R_T} rdr\,\int_0^{2\pi}L(\phi,r)T_{AA}(r,b=0)d\phi}
{\int_0^{R_T} rdr\,\int_0^{2\pi}T_{AA}(r,b=0)d\phi}  
\ee
where $\langle L \rangle \sim 5.8 (6.2) fm $ for RHIC (LHC).
The nuclear modification factor, $R_{AA}$ is defined as
\bea
R_{AA}(p_T)
= \frac{\frac{dN_{AA}^{\pi^0(\eta)}}{d^2p_Tdy}}
{\left[\frac{dN_{AA}^{\pi^0(\eta)}}{d^2p_Tdy}\right]_0}
\label{raa}
\eea
where the suffix `0' in the denominator indicates that energy loss
has not been considered while evaluating the expression.
\begin{figure}[h]
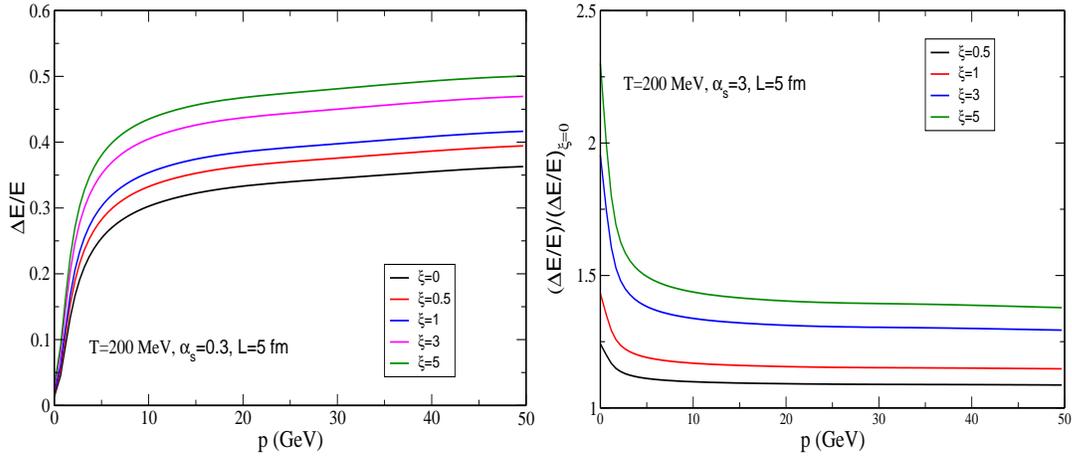

\begin{center}
\epsfig{file=debye_lq_T200_new.eps,width=7cm,height=6cm,angle=0}
\epsfig{file=rat_lq_T200_new.eps,width=7cm,height=6cm,angle=0}
\end{center}
\caption{(Color online) Fractional energy loss for light quark when mean free
path is independent of $\xi$ (left panel). The ratio of the fractional
energy loss of anisotropic media to that in isotropic media is also presented
(right panel)
}
\label{fig1}
\end{figure}

\subsection{Space-time evolution}
For an expanding plasma the anisotropy parameter $\xi$ and the hard momentum
scale $p_{\rm hard}$ (appearing in Eq.(\ref{debye}) via $\lambda$ as well
as in the expression of the Debye mass) are time
dependent.
Thus to calculate $R_{\rm AA}$ 
one needs to know the time dependence of $p_{\rm hard}$ and $\xi$. 
We shall follow the work of Ref.~\cite{mauricio} to evaluate the
$p_T$ distribution of hadrons from the first few Fermi of the 
plasma evolution. Three scenarios of the space-time evolution (as 
described in Ref.~\cite{mauricio}) are the following:
(i) $\tau_{\rm iso} = \tau_i$, the system evolves hydrodynamically
so that $\xi =0$ and $p_{\rm hard}$ can be identified with the
temperature ($T$) of the system (till date all the calculations have been
performed in this scenario), (ii) $\tau_{\rm iso}\rightarrow \infty$,
the system never comes to equilibrium, 
(iii) $\tau_{\rm iso} > \tau_i$ and $\tau_{\rm iso}$ is finite, one should
devise a time evolution model for $\xi $ and $p_{\rm hard}$ which smoothly 
interpolates between pre-equilibrium anisotropy and hydrodynamics. We 
shall follow scenario (iii) (see Ref.~\cite{mauricio} for details) 
in which case the time dependence of the anisotropy parameter $\xi$ is 
given by
\begin{eqnarray}
\xi(\tau,\delta) &=& \left(\frac{\tau}{\tau_i}\right)^\delta-1 
\label{eq_xi}
\end{eqnarray}
where the exponent $\delta = 2~(2/3)$ corresponds to {\em free-streaming 
(collisionally-broadened)} pre-equilibrium momentum space anisotropy and
$\delta=0$ corresponds to thermal equilibrium. 
As in Ref.~\cite{mauricio}, a transition width $\gamma^{-1}$ is introduced 
to take into account the smooth transition from non-zero value of $\delta$ 
to $\delta = 0$ at $\tau = \tau_{\rm iso}$. The time dependence of 
various quantities are, therefore, obtained in terms of a smeared step 
function \cite{prl100}:
\begin{equation} 
\Lambda(\tau)=\frac{1}{2}(\tanh[\gamma(\tau-\tau_{\rm iso})/\tau_{\rm iso}]+1). 
\label{eq_lamda}
\end{equation}

\begin{figure}[t]
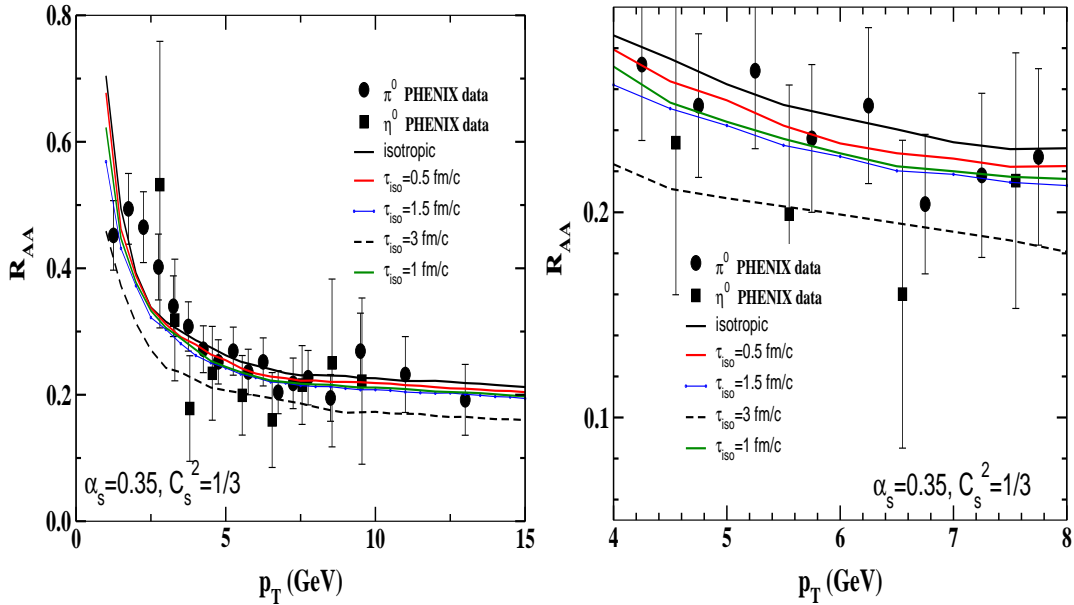

\begin{center}
\epsfig{file=raa_T440_cs13_als35.eps,width=7cm,height=8cm}
\epsfig{file=2raa_T440_cs13_als35.eps,width=7cm,height=8cm}
\caption{(Color Online) Nuclear modification factor at RHIC energies. The initial
conditions are taken as $T_i$=440 MeV and $\tau_i$=0.147 fm/c. Left
panel represents the results in the full $p_T$ range whereas the in the
right panel the results are given for 4 $ \leq p_T \leq $ 8 GeV }
\label{fig2}
\end{center}
\end{figure}

For $\tau << \tau_{\rm iso} ( >> \tau_{\rm iso})$ we have $\Lambda = 0 (1)$
which corresponds to {\em free streaming} (hydrodynamics). With this, the 
time dependence of relevant quantities are as follows~\cite{mauricio}:
\begin{eqnarray}
\xi(\tau,\delta) &=& \left(\frac{\tau}{\tau_i}\right)^
{\delta(1-\Lambda(\tau))}-1,\nonumber\\
p_{\rm hard}(\tau)&=&T_i~{\bar {\cal U}}^{c_s^2}(\tau),
\label{xirho}
\end{eqnarray}
where, 
\begin{eqnarray}
{\mathcal U}(\tau)&\equiv& \left[{\mathcal
    R}\left((\frac{\tau_{\rm iso}}{\tau})^\delta-1\right)
\right]^{3\Lambda(\tau)/4}
\left(\frac{\tau_{\rm iso}}{\tau}\right)^{1-\delta(1-\Lambda(\tau))/2},
\nonumber\\
{\bar {\cal U}}&\equiv& \frac{{\cal U}(\tau)}{{\cal U}(\tau_i)},
\label{utau}
\end{eqnarray}
$T_i$ is the initial temperature of the plasma and $c_s^2$ is the velocity of
sound. 
For isotropic case, we have $p_{\rm hard} = T, \tau_{\rm iso}=\tau_i$ 
so that $\Lambda = 1$, ${\cal U}(\tau)=\tau_i/\tau$, and ${\cal U}(\tau_i)=1$. 
By using $c_s^2 =1/3$ we recover
the Bjorken cooling law~\cite{bj}.
We assume here that the plasma expands longitudinally as the effect of
transverse expansion at the very early stage might be neglected. This has
been illustrated in case of photon production in Ref.~\cite{renk}. Since
the momentum space anisotropy is an early stage phenomenon, this assumption
is justified. Even if the transverse expansion is important in the
very early stage, it will have two effects on the parton energy
loss : (i) The expanding geometry will increase the duration of propagation,
and (ii) the same expansion will cause the parton density to fall along
its path. These two effects partially compensate each other and the energy
loss is almost the same as in the case without the transverse 
expansion~\cite{vitev}. 

Because the colliding nuclei do have a transverse density profile,
we assume that the initial temperature profile is given by~\cite{moore}
\be
T_i(r) = T_i\,\left[2\left(1-r^2/R_T^2\right)\right]^{1/4}
\label{init}
\ee
Using Eqs.(\ref{xirho}) and (\ref{init}) we obtain the profile
of the hard momentum scale as
\be
p_{\rm hard}(\tau,r) = T_i\,\left[2\left(1-r^2/R_T^2\right)\right]^{1/4} 
{\bar {\cal U}}^{c_s^2}(\tau)
\ee

In the present work it is assumed that an isotropic QGP is formed
at an initial temperature $T_i$ and initial time $\tau_i$. Subsequent
rapid longitudinal expansion leads to an anisotropic QGP which lasts
till $\tau_{\rm iso}$.
Now in order to estimate the initial temperature we proceed as follows.
In case of isentropic expansion the experimentally measured hadron multiplicity
can be related to the initial temperature and thermalization time
by the following equation~\cite{hwa}:
\be
T_i^3(b_m)\tau_i=\frac{2\pi^4}{45\zeta(3)\pi\,R_T^2 4a_k}
{\Big <}\frac{dN}{dy}(b_m){\Big >}
\label{intemp}
\ee
where ${\Big <} dN/dy(b_m){\Big >}$ is the hadron (predominantly pions)
multiplicity
for a given centrality class with maximum impact parameter $b_m$,
$R_T$ is the transverse dimension of the system,
$\tau_i$ is the initial thermalization time,
$\zeta(3)$ is the Riemann zeta function and
$a_k=({\pi^2}/{90})\,g_k$ is the degeneracy of the system created,
where $g_k = (7/8 \cdot 2 \cdot 2 \cdot N_F \cdot N_c +2\cdot 8)$,
and $N_c$ being the number of colors.
The hadron multiplicity  resulting from $Au + Au$ collisions
is related to that from pp collision at a given
impact parameter and collision
energy by
\be
{\Big <} \frac{dN}{dy}(b_m){\Big >}=\left[(1-x){\Big <} N_{part}(b_m){\Big >}/2
+x{\Big <} N_{coll}(b_m){\Big >}\right]\frac{dN_{pp}}{dy}
\label{dndy}
\ee
where $x$ is the fraction of hard collisions.
$\langle N_{part}\rangle$
is the average number of participants and  $\langle N_{coll}\rangle$ is the
average number of collisions
evaluated by using Glauber model.
$dN_{pp}^{ch}/dy= 2.5-0.25ln(s)+0.023ln^2s$
is the multiplicity of the produced hadrons
in $pp$ collisions at centre of mass energy, $\sqrt{s}~\cite{KN}$.
We have assumed that $20\%$ hard (i.e. $x=0.20$ )and
$80\%$ soft collisions are responsible for initial entropy production. This
gives the desired multiplicity measured at RHIC energies.
For 0 - 10\% centrality (relevant for our case) we obtain $T_i$ =
440 (350) MeV for $\tau_i = $ 0.147 (0.24) fm/c.

We also calculate the nuclear modification factor at LHC energies 
at $\sqrt{s}=2.76$ TeV and $\sqrt{s}=5.5$ TeV.
For the former case, the initial conditions are estimated as follows.
The measured charged particle multiplicity density at this energy is 
\cite{alice1} $dN_{\rm ch}/dy=1600$ in most central collisions corresponding
to $\langle N_{\rm part} \rangle \sim 382 $ and 
$\langle N_{\rm coll} \rangle \sim 1700$~\cite{alice2}.
Using Eq.(\ref{intemp}) we obtain $T_i \sim 650 $ MeV with
$\tau_i$ = 1 fm/c.
For LHC energies at $\sqrt{s}=5.5$, the multiplicity in 
$p-p$ collision is not known yet. We use the empirical formula for 
$dN_{pp}/dy 
\sim 0.8\ln \sqrt{s}$ which yields $T_i$=830 
for $\tau_i=0.08$
fm/c.


\section{Results}

For the quantitative estimates of the fractional energy loss
in an anisotropic media, we, first consider a plasma at a temperature
$T$ = 200 MeV with effective number of degrees of freedom
$N_F$=2.5 with the strong coupling as constant $\alpha_s$=0.3.
The fractional energy loss for non-zero $\xi $ ($\xi$=0.5, 1, 3, 5)
for light flavour is shown in Fig.~(\ref{fig1}) when the quark
propagates along the direction of anisotropy. It is observed that as
the anisotropy parameter ($\xi$) increases, the fractional energy loss
increases. The enhancement factor can be better understood by looking
at the right panel of Fig.~(\ref{fig1}) where we have plotted the ratio of the fractional
energy loss in anisotropic media to that in isotropic case.
It is seen that at low momentum the enhancement is more and after that it 
saturates for all the values of $\xi$ considered here.
For larger values of the anisotropic parameter, the ratio is seen
to increase reaching a maximum value of the order of 2 corresponding 
to $\xi=5$. It is to be noted that the fractional energy loss decreases
when the direction of propagation is not aligned with the anisotropy
direction~\cite{prabhee}. 

      It is observed that the energy loss increases with the 
anisotropy parameter $\xi$.
      Mathematically, this can be understood as follows. The energy loss
      is proportional to the square of the two-body potential $v({\bf q},\xi)$
      which has a factor $R(\xi)$ in the denominator. Now this quantity
     decreases with $\xi$ and hence the fractional energy loss increases
     with $\xi$. It is also to be noted that the two-body potential
     is stronger when the parton propagates in the anisotropy direction and
     it decreases away from the anisotropy direction leading to less
     energy loss~\cite{prabhee,dumitru08,dumitru}.


Now let us turn to the calculation of nuclear modification factor
at RHIC energies. For the time evolution of the system Eqs.(31) and (34) are used
for free-streaming interpolating model ($\delta=2$). The initial conditions
are taken as $T_i$ = 440 MeV and $\tau_i$=0.147 fm/c. The results for
$R_{\rm AA}$ for
various values of the isotropization time ,$\tau_{\rm iso}$ have
been compared with the PHENIX data~\cite{nuclex06} in Fig.~(\ref{fig2}). 
It is quite clear from right panel of Fig.~(\ref{fig2}) that the value 
of $R_{AA}$
for anisotropic media is lower than that for the isotropic
media as the energy loss in the former is higher by a factor of 1.2 - 2
(see Fig.~(\ref{fig1})). It is also observed that as $\tau_{\rm iso}$ 
increases the value of $R_{\rm AA}$ decreases compared to its
isotropic value. This is because the hard scale $p_{\rm hard}$
decreases slowly as compared to the isotropic case, i. e. the cooling
is slow. Also we have checked that as $\tau_{\rm iso}$ increases
the rate of cooling becomes slower leading to larger value of the energy
loss. For reasonable choices of $\tau_{\rm iso}$, the
experimental data is well described. It is seen that increasing the
value of $\tau_{\rm iso}$ beyond 1.5 fm/c grossly underpredict the data.
We find that the extracted value of isotropization
time lies in the range $0.5 \leq \tau_{\rm iso} \leq 1.5$ fm/c. This is in
agreement with the earlier finding of $\tau_{\rm iso}$ using PHENIX
photon data~\cite{LB3}.

In order to see the sensitivity on the initial conditions we now consider
another set of initial conditions, $T_i$ = 350 MeV and $\tau_i$ = 0.24 fm/c. 
\begin{figure}[t]
\begin{center}
\epsfig{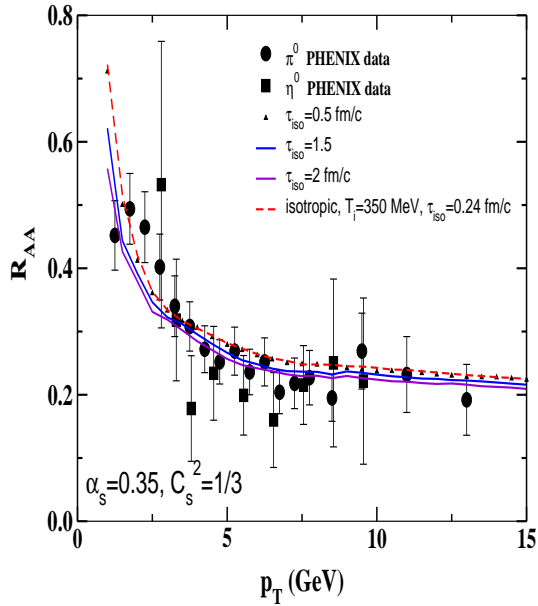}
\caption{(Color Online) Same as Fig.(\protect\ref{fig2}) at $T_i$ = 350 MeV and 
$\tau_i$ = 0.24 fm/c.}
\label{fig3}
\end{center}
\end{figure}
The result is shown in Fig.~(\ref{fig3}). In this case also the data is
well reproduced for the values of $\tau_{\rm iso}$ considered. It is
observed that to reproduce the data larger value of $\tau_{\rm iso}$ is
needed as compared to the case of higher initial temperature. We
extract an upper limit of $\tau_{\rm iso}$ = 2 fm/c in this case. 

\begin{figure}[t]
\begin{center}
\epsfig{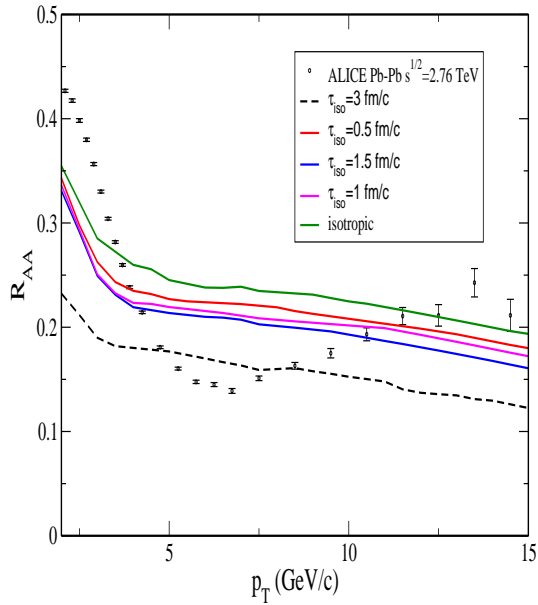}
\caption{(Color Online) Same as Fig.(\protect\ref{fig2}) at $\sqrt{s}=2.76$
TeV corresponding to $T_i$ = 650 MeV and 
$\tau_i$ = 0.1 fm/c.} 
\label{fig4a}
\end{center}
\end{figure}

We also compare our results to the recent measurement of $R_{AA}$ 
by the ALICE collaboration at $\sqrt{s}=2.76$ TeV~\cite{alice2}.
The result is shown in Fig.~(\ref{fig4a}). Note that LHC data, 
at this energy, show quite different behavior at high $p_T$(increasing trend)
 for which RHIC data do not exit. The reason might be due to larger volume and
 larger density. 
\begin{figure}[t]
\begin{center}
\epsfig{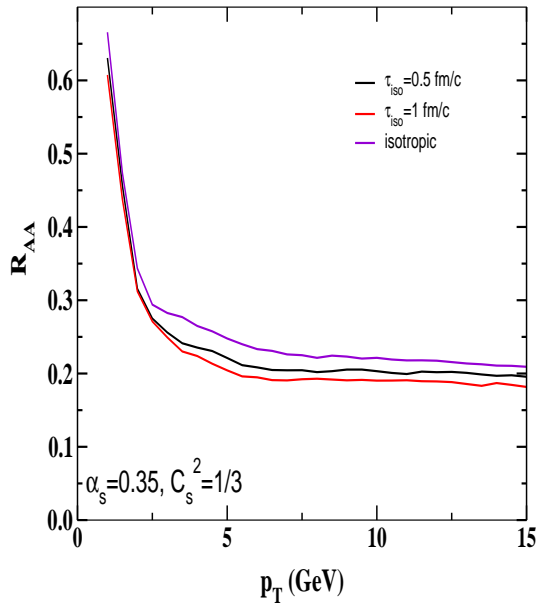}
\caption{(Color Online) Same as Fig.(\protect\ref{fig3}) at LHC energies with $T_i$ = 
830 MeV and 
$\tau_i$ = 0.08 fm/c.}

\label{fig4}
\end{center}
\end{figure}
However, unlike the RHIC data, 
our present model is unable to explain the LHC data (only 2-3 points
can be explained with the values of $\tau_{\rm iso}$ considered here).
Thus, within the ambit of the model used here, it is not possible to 
extract $\tau_{\rm iso}$ accurately at LHC energies. 
It is therefore, necessary to modify 
the present approach. The possible modifications might be the inclusions 
of collisional energy loss, path length distribution.
 
From the figures for $R_{AA}$ (both for RHIC and LHC
energies), it is observed
that the nuclear modification factor shows some kind of
saturation behavior with $\tau_{\rm iso}$. We have checked that
this occurs until $\tau_{\rm iso}$ = 1.5 and 2 fm/c for two
sets of initial conditions used at RHIC. Further increase
of $\tau_{\rm iso}$ leads to more suppression thereby
underpredicting the data (see Figs.~(\ref{fig2})) . Similar
observations have been noted while calculating photons from
anisotropic media~\cite{LB3}. 

\begin{figure}[t]
\begin{center}
\epsfig{file=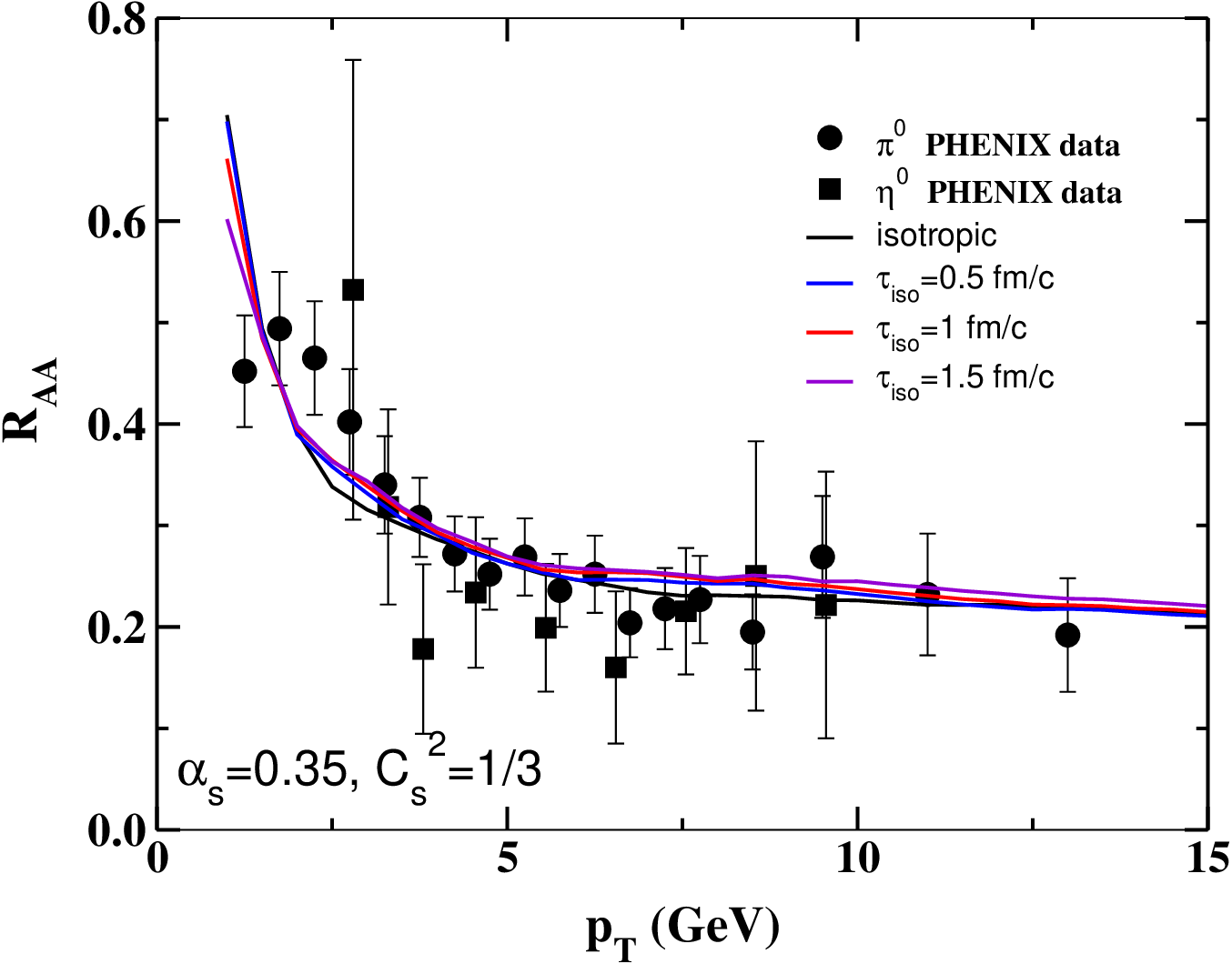,width=7cm,height=8cm}
\caption{(Color Online) Same as Fig.(\protect\ref{fig2}) for fixed
final multiplicity.}
\label{fig5}
\end{center}
\end{figure}

Although for RHIC data we could extract $\tau_{\rm iso}$, 
the extraction of $\tau_{\rm iso}$ is uncertain
by comparing with the LHC data at $\sqrt{s}=2.76$ TeV. However,
for the sake of completeness, we predict
the nuclear modification factors at higher LHC energies, i. e. 
at $\sqrt{s}=5.5$ TeV.
Since the final multiplicity at LHC energies is higher the initial temperature
is higher compared to RHIC energies. Also the life time of the 
plasma is large. Moreover, the length traversed by the jets in the
plasma is higher. As a result, one would expect that the energy loss
would be  more. Consequently, $R_{\rm AA}$ will be lower for the same values
of isotropization time, $\tau_{\rm iso}$. The results for  
various $\tau_{\rm iso}$ is delineated in Fig~(\ref{fig4}). As discussed
earlier, we find that the values of the nuclear modification factors
are less as compared to the case for RHIC. 

So far we have used the interpolating models which
assumes fixed initial conditions. However, enforcing fixed initial
condition results in generation of
particle number and enhance the final multiplicity during the
transition from $\delta = 2$ to zero. Because most of the experimental
results correspond to fixed final multiplicity, we should device
a mechanism which enforces fixed final multiplicity. To do this
the initial conditions have to be varied with the isotropization time, i. e.
one must lower the initial "temperature' for finite $\tau_{\rm iso}$. 
To ensure fixed final multiplicity in
this model one has to redefine $\bar{\cal U}(\tau)$ in
Eq. (31) as in ~\cite{mauricio}:
$$ 
\bar{\cal U}(\tau)= {\cal U}(\tau)\left[{\cal R}
((\tau_{iso}/\tau_i)^{\delta}-1)\right]^{-3/4}(\tau_{i}/\tau_{iso})
$$
This redefinition corresponds to a lower initial hard momentum scale
(${\em p_{hard}(\tau_i)}<T_i$) for $\tau_{iso}>\tau_i$. Larger value
of isotropization time corresponds to lower initial hard momentum scale.
As we shall wee, this has important consequences on the values
of $R_{AA}$.

The result for fixed final multtiplicity condition at RHIC energies has
been displayed in Fig.(\ref{fig5}). It is observed that the value
of $R_{AA}$ increases with $\tau_{\rm iso}$ compared to the isotropic case.
The reason for this is that in FFM the larger the value of $\tau_{\rm iso}$
is,
the lower is the initial hard momentum scale resulting less energy loss.
This gives rise to higher value of $R_{AA}$ as comapared to the fixed
initial condition (FIC) case as well as isotropic case. It is also seen that 
for FFM isotropic value of the nuclear modification factor is closer to the 
data in contrast to the
observation made in case of FIC where it is possible to extract the value
of the isotropization time at least for RHIC energies. Thus, use of FFM
does not lead to any firm conclusion about the extraction of $\tau_{\rm iso}$.
We have checked that increasing the value of $\tau_{\rm iso}$ beyond 1.5 fm/c
$R_{AA}$ increases grossly underpredicting the data.

\begin{figure}[t]
\begin{center}
\epsfig{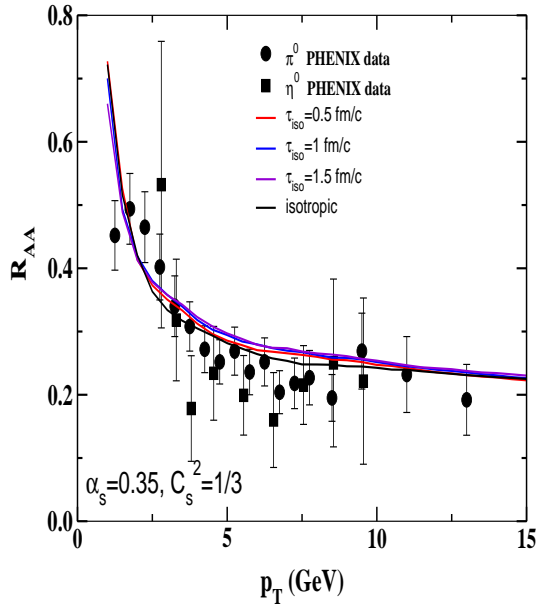}
\caption{(Color Online) Same as Fig.(\protect\ref{fig3}) for fixed
final multiplicity.}
\label{fig6}
\end{center}
\end{figure}

We reperform our calculation corresponding to Fig.(3) for FFM and the result is
shown in Fig.(\ref{fig6}). Similar behaviour (as in Fig.(\ref{fig5})) is
observed for the reason mentioned above.
Similarly, the calculations at LHC energies can be performed using FFM model.
It is easy to guess that
$R_{AA}$ will be larger as compared to the case with FIC model.

\section{Summary}

In this work, we have calculated the fractional energy loss
due to gluon radiation in an infinite size anisotropic media treating 
the scatterer as providing 
a screened coulomb-like potential. It is shown that the
presence of initial state anisotropy increases the radiative
energy loss by a factor of 1.2 - 2 depending upon the values of
the anisotropy parameter $\xi$ and the momentum of the jet when the parton
propagates along the direction of anisotropy.
We then calculate the hadron $p_T$ spectrum including this effect.
A phenomenological model for the space time evolution of
the anisotropic media has been used to obtain the time dependence of
the anisotropy parameter and the hard momentum scale.
The results for the nuclear modification factor for various
values of the isotropization time $\tau_{\rm iso}$ are then compared with
the PHENIX data. It is found that for FIC the data is well reproduced if
$\tau_{\rm iso}$ lies in the range $0.5 \leq \tau_{\rm iso} \leq 1.5$ fm/c 
which is in agreement with the previous findings. Increasing
$\tau_{\rm iso}$ beyond this range seems to underpredict the data.
 However, for FFM due to the lowering of the initial hard momentum
scale with $\tau_{\rm iso}$, $R_{AA}$ increases as compared to the case 
when FIC is used.
By comparing with the data it is concluded that it is very
difficult to infer which phase (isotropic or anisotropic) is favourable.

To cover the uncertainties
in the initial conditions we also consider another set of initial conditions
corresponding to RHIC energies.
It is observed that for lower initial temperature the upper limit of the
extracted value of $\tau_{\rm iso}$ is slightly higher as compared to the
case where larger value of initial temperature is used.
It is well known that the velocity of sound greatly influences the expansion
dynamics. 
It has been seen that the results are extremely sensitive to the 
velocity of sound.

After fixing the model parameters from RHIC data, we calculate the nuclear
modification factor at LHC energies both at $\sqrt{s}=2.76 $ and 5.5 TeV. 
The data at $\sqrt{s}=2.76$ is compared with our calculation.
It is to be noted that the data at $\sqrt{s}=2.76 $ show quite different
behavior (decreasing between $p_T=5 - 10$ GeV and again increasing)
than at RHIC energies. It is shown that the present model is unable to predict 
the isotropization time at these energies. In order to do that further 
refinement of the model is necessary.
Note that as the initial temperature is
higher, the nuclear modification factors are lower as compared to the
RHIC case for the same set of parameters. 

It is to be noted that we have not included the collisional energy loss
in the present work as its contribution will be subleading. However,
a complete calculation must include both the energy loss mechanisms in
order to estimate $\tau_{\rm iso}$ i. e. better constrain can be
imposed on $\tau_{\rm iso}$ in such case.

It is worthwhile to mention that the path length fluctuation is 
an important phenomenon
as shown in Ref.~\cite{wicksnpa} while explaining the non-photonic
single electron data. In addition to the anisotropy parameter $\xi$,
the energy loss of parton in anisotropic media also depends on the
direction of propagation with respect to the anisotropy axis. Therefore,
consideration of fluctuating
path length might lead to additional direction dependence. 
In addition, use of path length distribution (instead of constant value)
leads to surface emission effects which 
plays significant role in determining $R_{AA}$.

It is to be mentioned that we have used the average energy loss to
calculate the nuclear modification factor. However, to improve
our calculation statistical treatment of the energy loss has
to be incorporated. 
In our calculation of energy loss
Landau-Pomeranchuk effect has been neglected which is worth investigating.

We also do
not consider the 
recoil of the scatterer in this work. However,
this condition can be relaxed by incorporating the recoil corrections
which plays an important role as shown in Ref.~\cite{magdalenaprc}. Also
this will lead to the unstable modes which greatly influences
the transport coefficients, such as the radiative energy loss. This
will be included in future publication. Furthermore,
the finite size effect to the radiative energy loss
in anisotropic media would also be interesting to study.
We leave these issues for now.

\end{document}